# Using Cognitive Dimensions Questionnaire to Evaluate the Usability of Security APIs

Chamila Wijayarathna    Nalin A. G. Arachchilage    Jill Slay
Australian Centre for Cyber Security
School of Engineering and Information Technology
University of New South Wales
c.diwelwattagamage@student.unsw.edu.au,   nalin.asanka@adfa.edu.au,   j.slay@adfa.edu.au

**Abstract**
Usability issues that exist in security APIs cause programmers to embed those security APIs incorrectly to the applications they develop. This results in introduction of security vulnerabilities to those applications. One of the main reasons for security APIs to be not usable is currently there is no proper method by which the usability issues of security APIs can be identified. We conducted a study to assess the effectiveness of the cognitive dimensions questionnaire based usability evaluation methodology in evaluating the usability of security APIs. We used a cognitive dimensions based generic questionnaire to collect feedback from programmers who participated in the study. Results revealed interesting facts about the prevailing usability issues in four commonly used security APIs and the capability of the methodology to identify those issues.

## 1. Introduction

Cyber-attacks and data breaches have become exceedingly common that we hear about such incidents more often than not (Acar, Fahl, & Mazurek, 2016). Even though new technologies and methodologies to secure applications and data are introduced, so far they have failed to reduce the number of attacks and data breaches (Acar et al., 2016). One of the main reasons for this failure is that some security mechanisms are not easily learnable and understandable to programmers, and they find it difficult to use those mechanisms correctly when developing applications (Wurster & van Oorschot, 2009). Specially, security Application Programming Interfaces (APIs) that provide the interface for programmers to access security mechanisms are often not usable (Acar et al., 2016) (Wurster & van Oorschot, 2009) (Myers & Stylos, 2016). Therefore, programmers may end up using them incorrectly and hence, results in introducing security vulnerabilities to the applications they develop.

If the usability of security APIs can be improved, it will help programmers develop more secure applications and therefore, will help prevent possible cyber-attacks and data breaches (Acar et al., 2016) (Myers & Stylos, 2016). However, so far security API developers have failed to develop security APIs in a way that programmers who are using security APIs will find them usable. One of the major obstacles for developing usable security APIs is the lack of a proper methodology to evaluate the usability of security APIs (Myers & Stylos, 2016) (Mindermann, 2016). If there is a proper methodology to evaluate the usability of a security API and identify usability issues that may exist in it before it is delivered as a finished product, developers can fix those issues and deliver a more usable product (Mindermann, 2016).

The main objective of our research is to facilitate API developers to develop usable security APIs, by proposing a systematic approach to evaluate the usability of security APIs. To achieve this objective, first we will assess the applicability of existing API usability evaluation techniques, which have been used to evaluate the usability of general APIs, for evaluating the usability of security APIs. In this particular study, we focus on the API usability evaluation methodology based on the cognitive dimensions questionnaire (Clarke, 2004). In our previous work, through a literature survey, we argued that the cognitive dimensions framework and questionnaire proposed by Clarke (2004) may not be sufficient to evaluate the usability of security APIs (Wijayarathna, Arachchilage, & Slay, 2017). Therefore, we proposed an enhanced version of this framework and questionnaire to be used in security API usability evaluations (Wijayarathna et al., 2017). In this study, we evaluate the applicability of both these questionnaires (i.e. Clarke (2004)'s questionnaire and our questionnaire (Wijayarathna et al., 2017)) for evaluating the usability of security APIs through an empirical study.

We are in the process of conducting an empirical study using participants who are programmers. Even though we have not completed the study, early results revealed useful insights that could be used to outline a methodology for usability evaluations of security APIs. This work in progress paper will discuss the study we conducted and the observations made so far.

## 2. Related Work

Several methods have been introduced to improve the usability of general APIs such as cognitive dimensions questionnaire based empirical evaluation (Clarke, 2004), heuristic evaluation (Grill, Polacek, & Tscheligi, 2012) and API walk through method (O'Callaghan, 2010). However, these methodologies do not consider security related usability characteristics such as "security knowledge prerequisites", "hard to misuse" and "amount of security related code programmer has to implement" when evaluating the usability of APIs (Green & Smith, 2016) (Wijayarathna et al., 2017). This suggests the inapplicability of these existing evaluation techniques for evaluating the usability of security APIs.

In this particular study, we are focusing on the cognitive dimensions questionnaire based API usability evaluation methodology (Clarke, 2004), because it seems to be more effective than the other methodologies for evaluating the usability of security APIs (Wijayarathna et al., 2017). In this methodology, experimenters recruit programmers and ask them to complete an individual task which may involve writing, reading or debugging a code that use the API under evaluation (Clarke, 2004). Once the task is completed, each participant has to answer the cognitive dimensions questionnaire based on their experience in completing the task. Evaluators will identify usability issues exist in the API by going through the answers for the questionnaire provided by the participant.

The cognitive dimensions framework and questionnaire used by Clarke (2004) consists of 12 dimensions and questions to cover each of these dimensions. However, these 12 dimensions does not cover security API related usability aspects such as "security knowledge prerequisites", "hard to misuse" and "amount of security related code programmer has to implement" (Wijayarathna et al., 2017). Therefore, in our previous work, we proposed an enhanced version of the framework and questionnaire that consists of 15 dimensions (Wijayarathna et al., 2017). New questionnaire we proposed consist of all the questions used by Clarke (2004) as well as new questions that we introduced. The work described in this paper intends to evaluate the applicability of both these versions of cognitive dimensions framework and questionnaire for evaluating the usability of security APIs.

## 3. Methodology

For this study, we used four different programming tasks that use four different security APIs. We used more than one API because, we are trying to derive a generalized result for all security APIs. The APIs we used are Google authentication API, Bouncy castle light weight crypto API, OWASP Enterprise Security API (ESAPI) and a proprietary API that provides SSL related functionalities. For each API, we designed a programming task which makes use of the most important objects and methods exposed by the API. For example, in the task which uses the Google authentication API, participants had to make use of signIn and signOut functions and other supporting functions. All the tasks were designed to use Java or JavaScript programming languages.

We recruited participants who had at least one year of experience as a programmer. At the recruitment, they had to complete a short demographic questionnaire where we collected information such as their experience as a programmer, their proficiency in Java and JavaScript programming languages and their previous experience with the aforementioned four security APIs. Then we assigned a programming task from the four tasks to each participant. They could complete it in our laboratory or remotely using their own computer. While completing the task, they had to verbalize their thoughts and their computer screens were recorded. On completion, they had to answer our proposed questionnaire (which consisted of all the questions used by Clarke (2004) as well as new questions that we added).

Then the first author analyzed the answers provided by each participant to the questionnaire and identified usability issues each participant came up with. Then he divided those issues as "identified by

Clarke (2004)'s questions" and "identified by questions we added". Then the first author went through the screen recordings, think aloud results and code artifacts produced by participants, and identified usability issues each participant came up with.

## 4. Results

So far we collected and analyzed data of 7 participants. 3 of them did the task which required to use OWASP ESAPI, 2 participants did the SSL task, 1 participant did the task which required to use Google authentication API and 1 participant did the task which required to use Bouncycastle crypto API. 5 participants completed the study remotely while 2 participants completed the study in laboratory. Every participant used their preferred Integrated Development Environment (IDE) where 5 participants used Intellij Idea, 1 participant used Netbeans and 1 participant used Eclipse.

Table 1 summarize the number of potential usability issues identified by each participant with each method and those numbers as a percentage of the total number of issues identified by the particular user.

| Participant number | Total number of issues identified | Number of issues identified through observation | Number of issues identified to the complete questionnaire | Number of issues identified from the answers to the Clarke's questions |
|---|---|---|---|---|
| 1 | 29 | 16 (56%) | 15 (52%) | 9 (31%) |
| 2 | 8 | 3 (38%) | 6 (75%) | 3 (38%) |
| 3 | 13 | 5 (38%) | 10 (85%) | 8 (62%) |
| 4 | 20 | 5 (25%) | 19 (95%) | 15 (75 %) |
| 5 | 12 | 1 (8%) | 11 (92%) | 9 (75%) |
| 6 | 17 | 12 (71%) | 9 (53%) | 5 (29%) |
| 7 | 15 | 7 (47%) | 10 (67%) | 7 (47%) |
| Mean Percentages | | 40% (sd = 20.5%) | 74% (sd = 17.6%) | 51% (sd = 19.7%) |

*Table 1 – Issues Identified by each participant in each method.*

## 5. Conclusion and Discussion

From the results we captured, there were few main conclusions that we could made. Average 74% (SD = 17.6%) from the total issues identified by each user have been revealed from his/her responses to the questionnaire. This suggests that questionnaire method is effective for collecting feedback from participants in API usability studies. However, the intersection between issues identified by observing users and issues identified through the questionnaire was small (mean = 25.8% , sd = 17.6%).

By observing 7 participants and analyzing the code they provided, we identified 44 potential usability issues that exist in the 4 APIs used. Out of these 44 issues, only 20 (45%) were revealed by the questionnaire answers. Even though the other 24 issues were not identified by the responses to the questionnaire, questionnaire answers gave a high-level idea about some of these issues. For example, by observing the participant who did the programming task that used Bouncycastle API, we identified that "parameters of Scrypt.generate() method are not obvious". This issue was not identified by the answers that participant provided for the questionnaire. However, his answers to the questionnaire mentioned that "API does not reveal information about function parameters and what they return" which gives a high-level opinion about the previously discussed issue. Giving a high-level view of the actual usability issue has advantages as well as disadvantages. The questionnaire answer does not directly reveal the exact place the issue exist. Therefore, evaluators will not be able to locate where the exact issue is. If the task only requires to use few objects and functions, evaluators will be able to identify where the issue is easily. However, receiving a high-level idea could be useful in some cases. For example, for large APIs like Bouncycastle, it is not practical to design a programming task that will cover all the objects and functions exposed by the API. However, as the questionnaire provides a general feedback on the usability issues, evaluators can use them as guidelines to identify usability issues that exist in the components of the API, including those that were not used in the programming task. Sub devices concept that was introduced by Blackwell and Green (2000) can possibly be the solution for the above mentioned issue. In API context, these sub devices can be mapped into classes/functions of the API. However, Clarke (2004)'s cognitive dimensions questionnaire has not used this and also this will lengthen the questionnaire.

Furthermore, questionnaire method identified some issues that could not be identified by observation and code analysis. Each participant's questionnaire answers revealed average 8.8 (SD=3.8) issues that could not be identified by observing the participant and by analyzing the code s/he provided. Specially, questionnaire method seems to reveal issues related to cognitive dimensions such as progressive evaluation, premature commitment, API elaboration, consistency, end user protection and testability that could not identify by observing the participant and analyzing the code artifacts provided by the participant.

Our improved version of questionnaire revealed 11.6 (sd = 4.6) usability issues per participant compared to Clarke (2004)'s questionnaire which revealed 8 (sd = 3.7) issues per participant. This suggests that the improved version of the cognitive dimensions questionnaire is more effective in identifying usability of security APIs compared to the questionnaire proposed by Clarke (2004).

## 6. Future Work

One of the main limitations of our current results is that the usability issues were identified solely by the first author. The identified issues may change if this analysis was done by a different person. This is not only a limitation of the study we conducted, but also a limitation when using cognitive dimensions based usability evaluation for identifying usability of APIs. One of the possible solutions for this is to do the analysis by several persons (Analyst Triangulation), so that we might get a more general result.

This is a still work in progress where we have only used 7 participants so far. We will continue doing this study with more participants to get a more statistically significant result on the applicability of cognitive dimensions questionnaire for security API usability evaluations.